# Significant enhancement of superconductivity under hydrostatic pressure in $CeO_{0.5}F_{0.5}BiS_2$ superconductor

Rajveer Jha, H. Kishan and V.P.S. Awana[*]

CSIR-National Physical Laboratory, Dr. K.S. Krishnan Marg, New Delhi-110012, India

We report the effect of hydrostatic pressure (0 – 1.97 GPa) on the superconductivity of $BiS_2$ based $CeO_{0.5}F_{0.5}BiS_2$ compound. The $CeO_{0.5}F_{0.5}BiS_2$ superconductor was synthesized by the solid state reaction route and the compound is crystallized in tetragonal P4/nmm space group. The studied compound shows superconductivity with transition temperature of 2.5K ($T_c^{onset}$) at ambient pressure, which has been enhanced to 8 K at applied pressure of 1.97 GPa. The observed normal resistivity exhibited semiconducting behavior. The data of normal state resistivity ρ(T) has been fitted by activation type equation and it is found that the energy gap is significantly reduced with pressure. Resistivity measurements under magnetic field for the highest applied pressure of 1.97GPa ($T_c^{onset}$ = 8K) exhibits the upper critical field of above 5Tesla. The observation of fourfold increase in $T_c$ accompanied with improved normal state conduction under hydrostatic pressure on $CeO_{0.5}F_{0.5}BiS_2$ superconductor calls for the attention of solid state physics community.

[*]**Corresponding Author**
Dr. V. P. S. Awana, Principal Scientist
E-mail: awana@mail.npindia.org: Ph. +91-11-45609357, Fax-+91-11-45609310
Homepage www.fteewebs.com/vpsawana/

Introduction

There has been a great interest in the scientific community after the discovery of $Bi_4O_4S_3$ superconductor with transition temperature $T_c$ = 8.6K [1,2] and $REO_{0.5}F_{0.5}BiS_2$ (RE=La, Ce, Nd and Pr) with $T_c$ = 2.5-10K [3-10]. The parent phase, $REOBiS_2$ (RE = La, Nd, Ce and Pr) exhibited bad metal behavior [3-10]. In these compounds the superconductivity is induced by the electron doping which is carried out via substitution of $F^{-1}$ at $O^{-2}$ site in the oxide block (REO). At the same time in similar structure i.e. $SrFBiS_2$ compound, the superconductivity has been observed via hole doping that is attained by the substitution of $La^{3+}$ at $Sr^{2+}$ site [11,12]. Therefore it can be safely said that superconductivity appears via carrier doping in the vicinity of the insulating-like state of the parent $BiS_2$ based compounds. Moreover, the superconductivity has also been induced via chemical substitution of the tetravalent ions such as $Th^{4+}$, $Hf^{4+}$, $Zr^{4+}$ and $Ti^{4+}$ in place of trivalent lanthanum $La^{3+}$, resulting in the increase of charge density in bad metal $LaOBiS_2$ compound [13]. In various $REO_{0.5}F_{0.5}BiS_2$ systems, the role played by the rare earth elements is noteworthy as for example replacing La with heavier rare earth elements, the $T_c$ (2.1K) of $LaO_{0.5}F_{0.5}BiS_2$ increases to 3.5K and 5K for $PrO_{0.5}F_{0.5}BiS_2$ and $NdO_{0.5}F_{0.5}BiS_2$ [3,4,10].

Both the experimental and theoretical studies of $BiS_2$ based superconductors have progressed well but the studies related to impact of external high pressure on the superconducting properties of this compound are limited to select groups only [14-16]. The application of external pressure may effectively change the superconducting properties such as transition temperature $T_c$ of these $BiS_2$ compounds. Also the same could effectively alter the lattice constants through bond lengths and angles that would certainly affect the electronic and magnetic correlations of a superconductor. In fact the $REO/FBiS_2$ samples being prepared under high pressure (HP) showed higher $T_c$ [3]. It has been speculated that increase in $T_c$ is due to a structural phase transition under pressure. In fact later it was observed by X-ray diffraction (XRD) measurements under ~1GPa pressure that a structural phase transition from tetragonal phase (P4/nmm) to monoclinic phase (P21/m) takes place in $LaO_{0.5}F_{0.5}BiS_2$ [17].

Keeping in view the importance of impact of the external pressure on the superconductivity of $BiS_2$ based superconductors we study the effect of hydrostatic pressure on the superconductivity of the $CeO_{0.5}F_{0.5}BiS_2$ compound having a $T_c$ of 2.2K at ambient pressure.

This compound is also reported to exhibit the coexistence of superconductivity and ferromagnetism either due to the possible local moments of the Ce or may be some unseen exotic reasons related to the superconducting mechanism [9,18]. Worth mentioning is the fact that impact of hydrostatic pressure on superconductivity of $CeO_{0.5}F_{0.5}BiS_2$ has already been reported very recently [14,15]. An increase in $T_c$ from 2.2K at ambient pressure to above 6K was observed at 1.64GPa, which remains nearly unaltered for studied higher pressures of up to 2.47GPa [14]. In the present report, the compound $CeO_{0.5}F_{0.5}BiS_2$ has been subjected to pressure dependent electrical resistivity in the temperature range 300 K down to 2K at several applied pressures of 0 to 1.97GPa to reproduce and test the earlier reported interesting result [14,15]. Our results are though qualitatively similar to that as reported earlier [14,15], yet quantitatively different. In our case the $T_c^{onset}$ increases to above 7K at just 0.97GPa. Also the superconducting transitions are relatively sharper in our case. We also did the magneto transport measurements under high pressure with maximum $T_c$ to estimate the upper critical field ($H_{c2}$) and compared the same with our reported ambient pressure results [18]. The normal state conduction of our studied compound is also seemingly better than the one being reported earlier [14,15]. This may be possible due to slight change in quality of the studied samples. In any case, without ambiguity the fourfold enhancement in $T_c$ at just 0.97GPa pressure calls for the attention of solid state physics community.

Experimental Details

We synthesized polycrystalline sample of $CeO_{0.5}F_{0.5}BiS_2$ by standard solid state reaction route via vacuum encapsulation. High purity Ce, Bi, S, $CeF_3$ and $CeO_2$ were weighed in stoichiometric ratio and ground thoroughly in a glove box under pure argon gas atmosphere. The mixed powder was subsequently palletized and vacuums sealed ($10^{-3}$Torr) in a quartz tube. Sealed quartz ampoule was placed in tube furnace and heat treated at $700^0C$ for 12hours with the typical heating rate of $2^0C$/minute followed by cooling down slowly over a span of six hours to ambient temperature and this entire process was repeated twice. X-ray diffraction (XRD) was performed at ambient temperature in the scattering angular range of $10^0 - 80^0$ (2θ) in equal step of $0.02^0$ using Rigaku Diffractometer with CuKα (λ=1.54Å). Rietveld analysis was carried out using the standard FullProf program. For carrying out measurement of pressure dependent electrical resistivity, HPC-33 piston type pressure cell compatible to DC resistivity Option of the Quantum Design's 14Tesla PPMS was used. The Hydrostatic pressures were generated by a BeCu/NiCrAl clamped piston-cylinder cell. The sample was immersed in a fluid pressure transmitting medium of Fluorinert (FC70:FC77=1:1) in a Teflon cell. Annealed Pt leads were affixed to gold-sputtered contact surfaces of the sample with silver epoxy in a standard four-wire configuration. The pressure at low temperature was calibrated from the superconducting transition temperature of Pb.

Results and discussion

Figure 1 shows the room temperature observed and Reitveld fitted XRD patterns of $CeO_{0.5}F_{0.5}BiS_2$ sample. The synthesized compound is well fitted in tetragonal structure with space group *P4/nmm*. The Rietveld refined lattice parameters are a = 4.037(2)Å and c = 13.407(3)Å. In the previous report [9] it has been shown that the c parameter for the parent phase $CeOBiS_2$ is 13.604Å, which is larger in comparison of $CeO_{0.5}F_{0.5}BiS_2$ sample. The decrease in c parameter of $CeO_{0.5}F_{0.5}BiS_2$ sample is a clear indication of $O^{-2}$ site $F^{-1}$ doping in parent $CeOBiS_2$ compound [18]. It is to be mentioned that the studied compound is contaminated with small impurity of Bi, which is marked with # in Fig. 1. The small Bi impurity appears close to the main peaks of $CeO_{0.5}F_{0.5}BiS_2$ tetragonal structure and is reportedly seen in most of $CeO_{1-x}F_xBiS_2$ samples [9]. We choose to work on x = 0.5 composition, because the $T_c$ is near maximum at this doping level [9]. Inset of the Fig 1 shows the resultant unit cell of the $CeO_{0.5}F_{0.5}BiS_2$ sample with *P4/nmm* space group. The layered structure is composed of CeO and $BiS_2$ layer along the *c* axis. Various atoms with their respective positions are indicated in crystal structure. Cerium (Ce), and Sulfur (S1and S2) atoms occupy the 2c (0.25, 0.25, z) site. On the other hand O/F atoms are located at 2a (0.75, 0.25, 0) site.

The temperature dependent electrical resistivity with applied pressures from 0GPa-1.97GPa for $CeO_{0.5}F_{0.5}BiS_2$ compound is shown in the Figure 2a. The electrical resistivity data is recorded from 300K down to 2K. Without applied pressure the normal state resistivity shows the semiconducting behavior down to the onset superconducting transition temperature $T_c^{onset}$. With an increase in applied pressure the normal state resistivity is though suppressed but yet exhibits semiconducting behavior. Figure 2(b) shows the expanded part of the Fig. 2(a) in the temperature range from 2K to 15K, to clearly visualize the superconducting transition temperature ($T_c$) of $CeO_{0.5}F_{0.5}BiS_2$ compound for the various applied pressures. At zero pressure $T_c(\rho = 0)$ is 2.2K and which is though nearly unchanged for 0.35GPa pressure, but the same increases slightly to 2.5K at 0.55GPa pressure. At 0.97GPa pressure the $T_c(\rho = 0)$ increases to above 3.5K with a broader transition width. In fact the $T_c^{onset}$ increases to above 7K at just 0.97GPa. It seems there is a sudden increase in onset of superconductivity at 0.97GPa. Both $T_c^{onset}$ and $T_c(\rho = 0)$ are marked in Figure 2b. At further higher pressure of 1.38GPa though the $T_c(\rho = 0)$ is increased sharply from 3.5K to 6K, the $T_c^{onset}$ increases to around 8K only. For still higher applied pressures of 1.68 and 1.97GPa the $T_c(\rho = 0)$ is seen at 7K with 8K onset $T_c$. To clearly visualize the pattern of increase in $T_c^{onset}$ and $T_c(\rho = 0)$, in Figure 2c we plot both the quantities against applied pressure for the studied $CeO_{0.5}F_{0.5}BiS_2$ compound. Seemingly though there is a sharp increase in both at 0.97GPa, but the $T_c^{onset}$ is enhanced relatively more. This gives rise to an increased broadening of transition at 0.97GPa, which is plotted in inset of Figure 2(c). The measurements are performed with increasing the pressure. Summarily it can be seen that the $T_c^{onset}$ initially increases up to 7K till 0.97GPa pressure and later saturates for the higher pressures of 1.97GPa. On the other hand, the normal state resistivity keep on decreasing with applied pressures up to 1.97GPa. The sharp increase in superconducting transition temperature under hydrostatic pressure of just around 1GPa may happen due to structural phase transformation being evident recently in these systems [17,19]. Also, there is a possibility of the presence of strong electron correlations in $BiS_2$ superconducting layers being enhanced under pressure.

To explore quantitatively the trend of change in normal state conduction with pressure, in Figure 3(a), we show the of $\log(\rho)$ vs $1/T$ plots for $CeO_{0.5}F_{0.5}BiS_2$ compound at various applied pressures. This is done to determine the values of energy gaps in semi-metallic regime. The observed data of the resistivity could be described in two different regions by the activation type relation $\rho(T) = \rho_0 e^{\Delta/2k_B T}$ where $\rho_0$ is a constant, $\Delta$ is the energy gap and $k_B$ is the Boltzmann constant. The observed resistivity versus temperature data could not be fitted for a single gap in entire temperature range. Hence the fitting has been done in two distinct regions of $\rho(T)$. Thus obtained energy gaps are marked as $\Delta_1$ in the temperature range 300 to 200K and $\Delta_2$ from 20K down to $T_c$ onset. The semiconducting behavior is suppressed with pressure in a similar fashion as observed earlier for another $BiS_2$ based superconductor $LaO_{0.5}F_{0.5}BiS_2$ [14, 15]. The analysis of $\rho(T)$ data at the ambient pressure gives the estimated values of the energy gaps $\Delta_1/k_B \approx$ 2024.55 K and $\Delta_2/k_B \approx 65.68$ K for the $CeO_{0.5}F_{0.5}BiS_2$ compound. Interestingly, the obtained values are considerably less than the ones reported by Wolowiec *et al.* for the $CeO_{0.5}F_{0.5}BiS_2$ compound [15]. May be the presently studied $CeO_{0.5}F_{0.5}BiS_2$ compound possesses improved normal state conduction than the one reported in ref. 15. Figure 3(b) shows the variation of calculated energy gaps ($\Delta_1/k_B$, $\Delta_2/k_B$) with various applied pressures from 0-1.97GPa. It can be clearly seen from Fig. 3(b) that both energy gaps $\Delta_1$ and $\Delta_2$ decrease rapidly with applied pressure of up to 1.68GPa and are almost saturated for 1.97GPa pressure. The decrease of energy gaps with increasing pressure, suggests the increase of charge carriers density at the Fermi surface with application of pressure. Interestingly, the first-principle calculations have suggested that there could be a insulator-metal transition accompanied with superconductivity at low temperature under pressure in these systems [20]. In the present study, we observed that up to 1.97GPa pressure though there is no insulator to metal transition, but the semiconducting behavior is suppressed significantly and superconductivity onset transition temperature is increased from around 2.5K to 8K. The interesting $BiS_2$ based $CeO_{0.5}F_{0.5}BiS_2$ superconductor need to be studied for further higher pressures.

Figure 4 represents the temperature dependent resistivity under applied magnetic field up to 5Tesla for $CeO_{0.5}F_{0.5}BiS_2$ sample at 1.97GPa pressure. The broadening in the superconducting transition $T_c$ with increasing applied field shows the typical type II superconductivity behavior for the $CeO_{0.5}F_{0.5}BiS_2$ compound. With applied magnetic fields both $T_c^{onset}$ and $T_c(\rho = 0)$ decrease to lower temperatures with an increase in broadening of the transition. The trend is similar to that as for high $T_c$ layered cuprate and Fe-pnictide superconductors. The upper critical field ($H_{c2}$) of $CeO_{0.5}F_{0.5}BiS_2$ sample at ambient and 1.97GPa hydrostatic pressure has been shown in the Figure 4b. The $H_{c2}$ is calculated from 90% criteria of normal state resistivity ($\rho_n$) by using the conventional one-band Werthamer–Helfand–Hohenberg (*WHH*) equation, i.e., $H_{c2}(0) = -0.693T_c(dH_{c2}/dT)_{T=Tc}$. The estimated values of $H_{c2}(0)$ are 1.5 Tesla for 0 pressure and 4.8 Tesla for 1.97GPa pressure. The $H_{c2}$ data points at ambient pressure have been taken from our earlier report on $CeO_{0.5}F_{0.5}BiS_2$ compound [18].

Conclusion

The phase pure tetragonal $CeO_{0.5}F_{0.5}BiS_2$ superconductor with $T_c$=2.2 K has been synthesized using solid state reaction route and studied the effect of hydrostatic pressure on its normal state and superconducting properties. The superconducting transition temperature of the synthesized compound is enhanced fourfold from around 2 K to 8 K at 1.97 GPa. The electrical resistivity data in the entire temperature range has been fitted using activation type equation revealing significant improvement in electrical conduction under pressure. The estimated value of the upper critical field for this compound at 1.97 GPa is above 5 Tesla. The observation of fourfold increase in $T_c$ accompanied with improved normal state conduction under just above 1GPa hydrostatic pressure on $CeO_{0.5}F_{0.5}BiS_2$ superconductor calls for the attention of solid state physics community.

Acknowledgement

Authors would like to thank their Director NPL India for his keen interest in the present work. This work is financially supported by *DAE-SRC* outstanding investigator award scheme on search for new superconductors. Rajveer Jha acknowledges the *CSIR* for the senior research fellowship. H. Kishan thanks CSIR for providing Emeritus Scientist Fellowship.

**Figure Captions**

**Figure 1:** Observed (*open circles*) and calculated (*solid lines*) XRD patterns of $CeO_{0.5}F_{0.5}BiS_2$ compound at room temperature. Inset is schematic unit cell of $CeO_{0.5}F_{0.5}BiS_2$ compound.

**Figure 2(a):** Resistivity versus temperature ($\rho$ Vs T) plots for $CeO_{0.5}F_{0.5}BiS_2$ compound, at various applied pressures in the temperature range 300K - 2.0K.

**Figure 2(b):** Resistivity versus temperature ($\rho$ Vs T) plots for $CeO_{0.5}F_{0.5}BiS_2$ compound, at various applied pressures in the temperature range 15K - 2.0K.

**Figure 2(c):** $T_c^{Onset}$ and $T_c(\rho = 0)$ versus pressure plots for $CeO_{0.5}F_{0.5}BiS_2$ compound, inset shows the variation of superconducting transition width with pressure. .

**Figure 3(a):** Resistivity ($\rho$) versus 1/T plot for up to 1.97 GPa pressure. The solid lines are linear fitting of activation energy equation at high and low temperatures.

**Figure 3(b):** Energy gaps $\Delta_1$ and $\Delta_2$ plotted as a function of pressure for $CeO_{0.5}F_{0.5}BiS_2$ compound.

**Figure 4(a):** Temperature dependence of the Resistivity $\rho(T)$ under magnetic fields for the $CeO_{0.5}F_{0.5}BiS_2$ compound under 1.97GPa hydrostatic pressure.

**Figure 4(b):** The upper critical field $H_{c2}$ taken from 90%, resistivity criterion $\rho(T)$ of $CeO_{0.5}F_{0.5}BiS_2$ compound at ambient and 1.97GPa hydrostatic pressure.

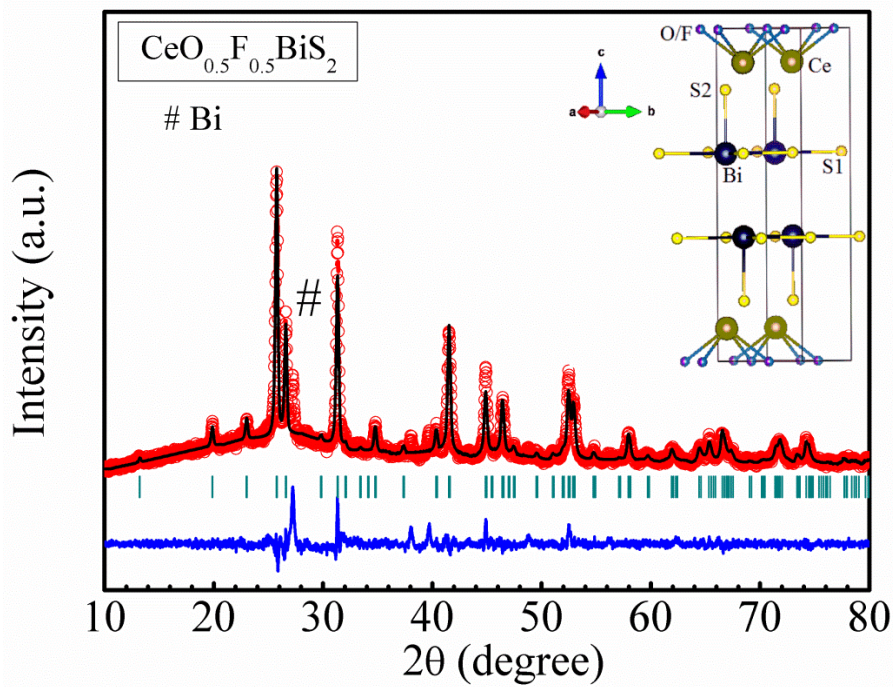

Figure 1

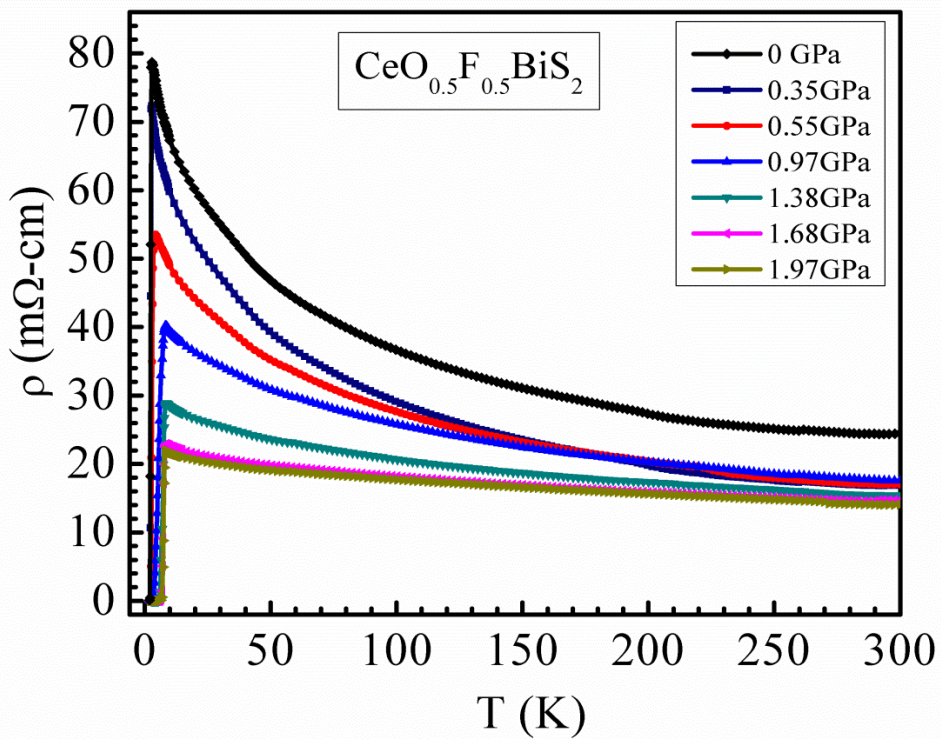

Figure 2(a)

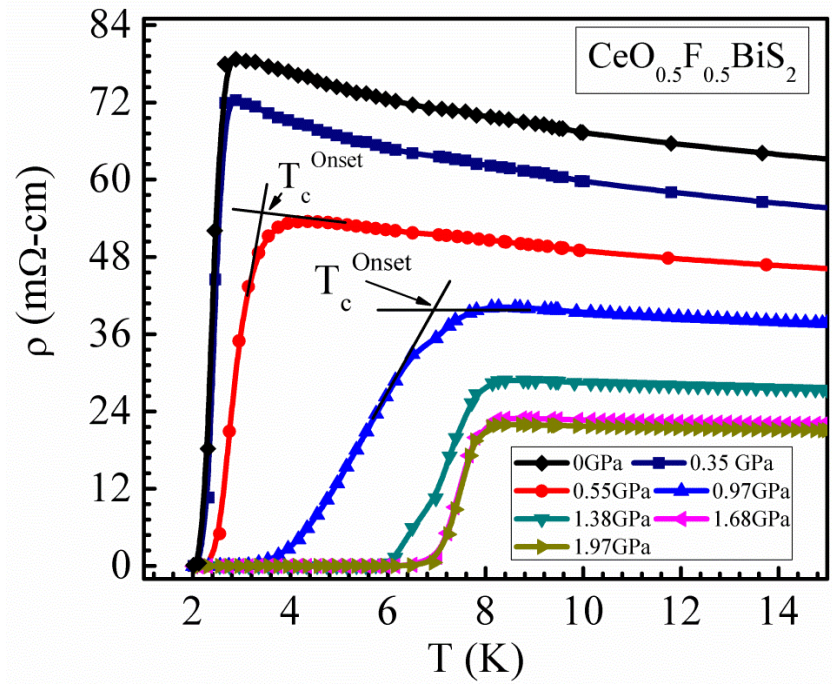

Figure 2(b)

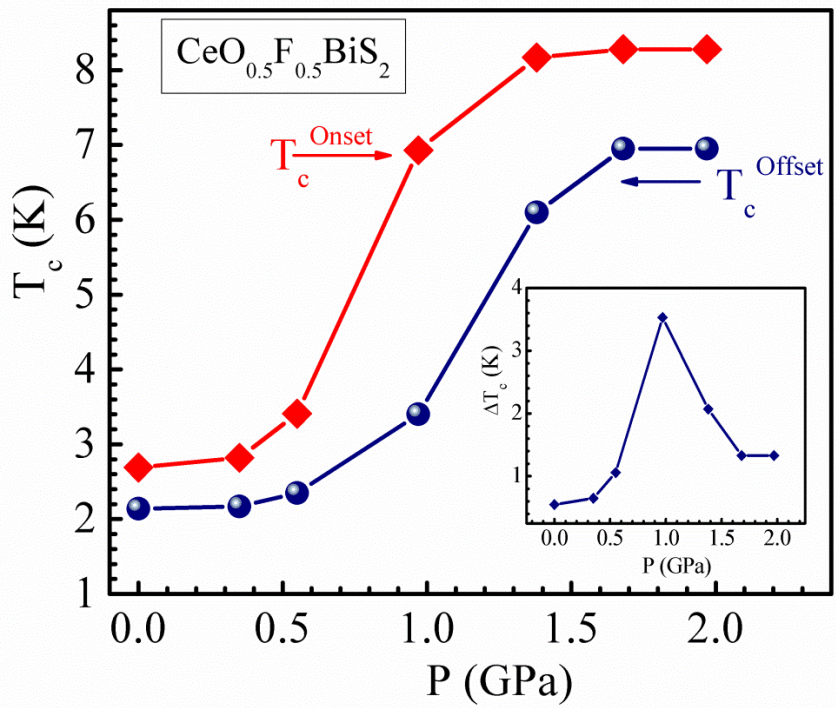

Figure 2(c)

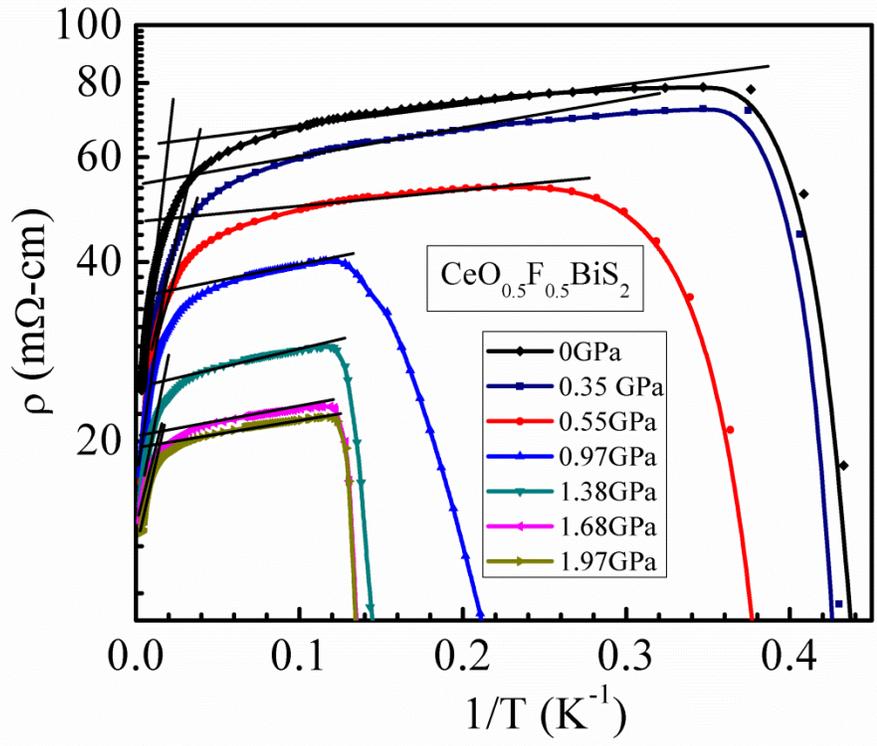

Figure 3(a)

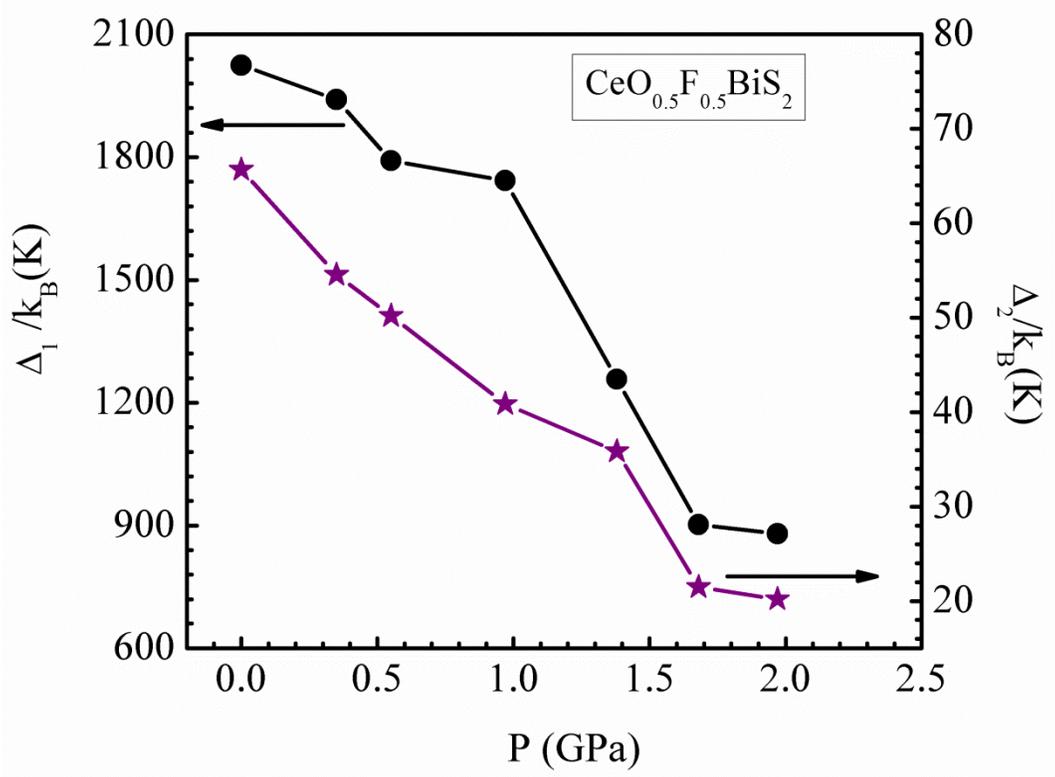

Figure 3(b)

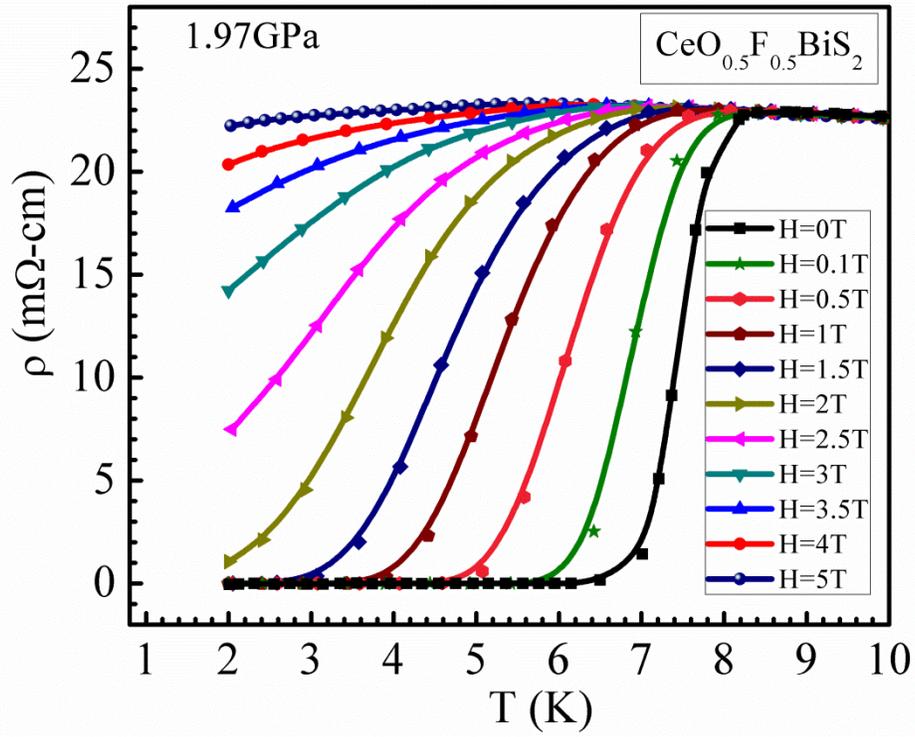

Figure 4(a)

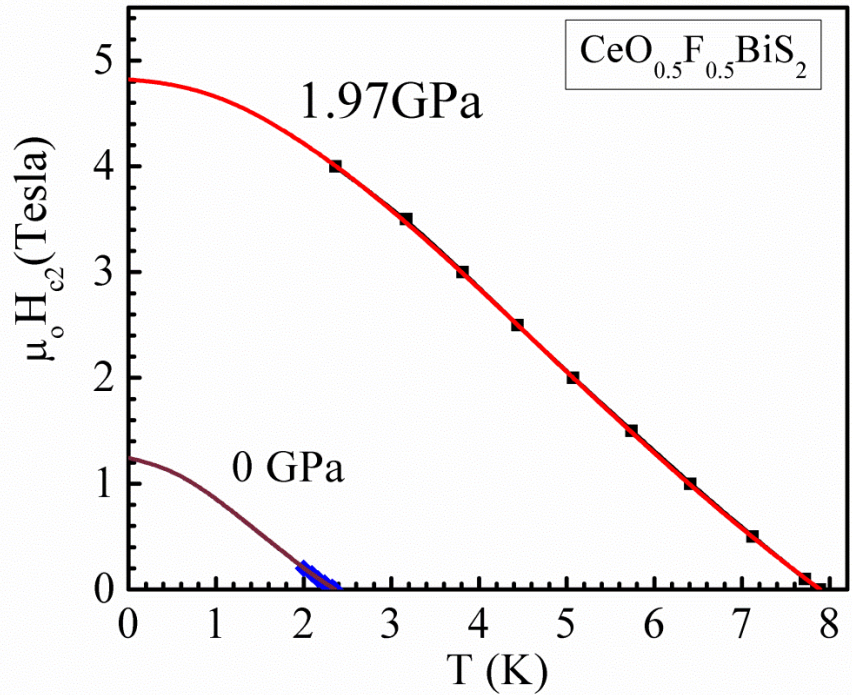

Figure 4(b)